\pdfoutput=1

\documentclass[11pt]{article}

\usepackage[final]{acl}

\usepackage{times}
\usepackage{latexsym}
\usepackage[T1]{fontenc}
\usepackage[utf8]{inputenc}
\usepackage{microtype}
\usepackage{inconsolata}
\usepackage{graphicx}

\usepackage{makecell}
\usepackage{booktabs}
\usepackage{multirow}
\usepackage{array}
\usepackage{svg}
\usepackage{float}
\usepackage{amsmath}
\usepackage{amssymb}    
\usepackage{enumitem}   
\usepackage{mathrsfs}   
\usepackage{tablefootnote} 

\usepackage{tikz}
 
\usepackage{hhline}
\usepackage{xcolor}
\usepackage{algorithm}
\usepackage{algorithmic}
\usepackage{amsfonts}

\usepackage{hyperref}
\usepackage{arydshln}
\usepackage{amssymb}
\usepackage{pifont}
\usepackage[export]{adjustbox}
\newcommand\blfootnote[1]{
    \begingroup
    \renewcommand\thefootnote{}\footnote{#1}
    \addtocounter{footnote}{-1}
    \endgroup
}
\title{gMBA: Expression Semantic Guided Mixed Boolean-Arithmetic Deobfuscation Using Transformer Architectures}

\author{Youjeong Noh,  Joon-Young Paik,  Jingun Kwon$^*$,  Eun-Sun Cho$^*$ \\
  Chungnam National University, Daejeon, Republic of Korea \\
  \small \texttt{youjoeng.noh14@o.cnu.ac.kr}, \space\space
  \small \texttt{\{lucadi, jingun.kwon, eschough\}@cnu.ac.kr}
}

\begin{document}
\maketitle

\begin{abstract}
Mixed Boolean-Arithmetic (MBA) obfuscation protects intellectual property by converting programs into forms that are more complex to analyze. However, MBA has been increasingly exploited by malware developers to evade detection and cause significant real-world problems. Traditional MBA deobfuscation methods often consider these expressions as part of a black box and overlook their internal semantic information. To bridge this gap, we propose a truth table, which is an automatically constructed semantic representation of an expression's behavior that does not rely on external resources. The truth table is a mathematical form that represents the output of expression for all possible combinations of input. We also propose a general and extensible guided MBA deobfuscation framework (gMBA) that modifies a Transformer-based neural encoder-decoder Seq2Seq architecture to incorporate this semantic guidance. Experimental results and in-depth analysis show that integrating expression semantics significantly improves performance and highlights the importance of internal semantic expressions in recovering obfuscated code to its original form.\blfootnote{\hspace{-4pt}$^*$Corresponding author}\footnote{Our code is available at https://github.com/Master-whiece/gMBA.} 
\end{abstract}

\section{Introduction}

Mixed Boolean-Arithmetic (MBA) obfuscation protects intellectual property by transforming code into complex forms that are difficult to analyze~\cite{nagra2009surreptitious,collberg2012distributed,ceccato2014need,bardin2017backward}. It combines Boolean (AND, OR, XOR) and arithmetic (+, -, ×) operations to create exponentially more complex expressions~\cite{zhou2007information}, making reverse engineering harder~\cite{10.1145/2382196.2382210}. However, these techniques are increasingly exploited by malware developers to evade detection. 
\setlength{\belowcaptionskip}{-8pt}
\begin{figure}[htb!]
    \centering
    \includegraphics[width=.88\linewidth]{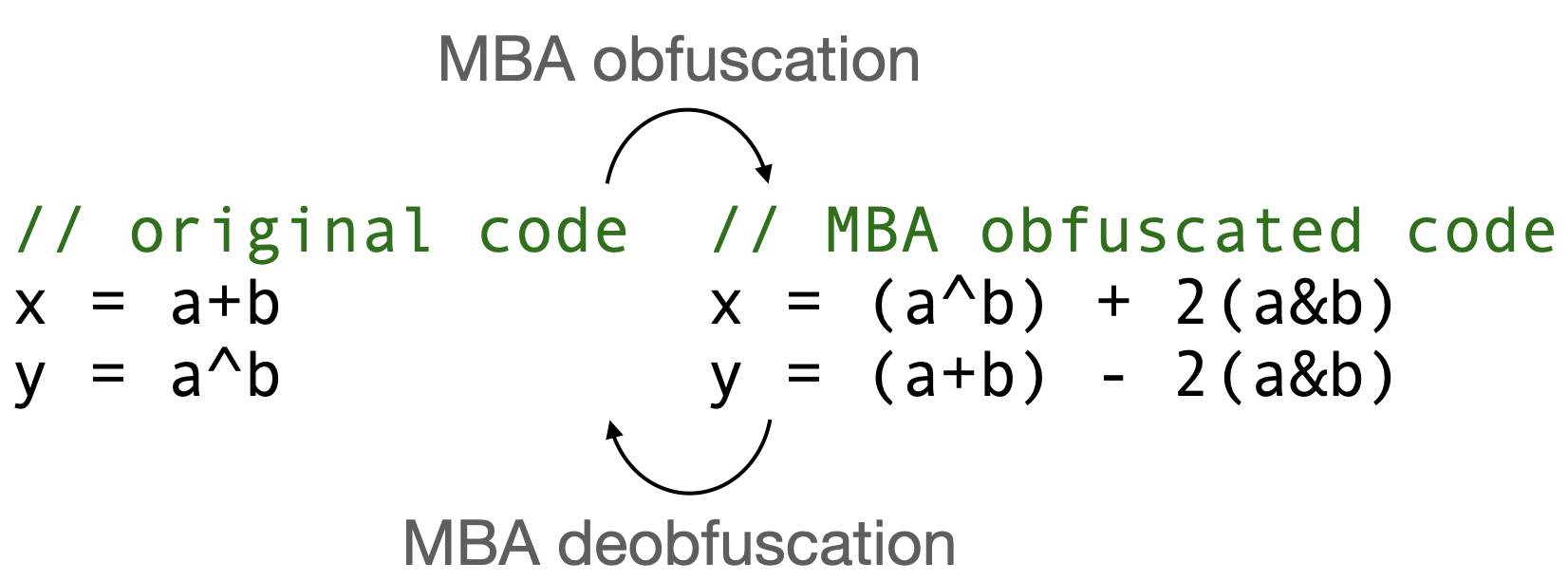}
    \caption{MBA obfuscation and deobfuscation transform mathematically equivalent expressions into different code representations.}
    \label{fig:enter-label}
\end{figure}
By obfuscating internal code, malware can bypass security analysis tools~\cite{cohen1987computer,kaur2024evolution}. 
For example, the DarkSide ransomware utilized obfuscation to evade detection~\cite{fbi_national_press_office_deputy_2021,trend_micro_research_what_2021}, highlighting the need for advanced deobfuscation techniques. 
Figure~\ref{fig:enter-label} shows an example of MBA obfuscation and deobfuscation.
 
Traditional MBA deobfuscation methods include bitblast~\cite{bitblast, liu2021mba}, pattern matching~\cite{10.1145/2995306.2995308, reichenwallner2022efficient}, and program synthesis~\cite{schloegel2022loki, 10.1145/3576915.3623186}. Recently, Seq2Seq models using RNNs \cite{rumelhart1986learning} and Transformers \cite{vaswani_attention_2023} have been introduced \cite{feng-etal-2020-neureduce}, but they consider these expressions as part of a black box and overlook their internal semantic information, leading to lower accuracy. To address this, we propose using \textit{truth tables} as a structured semantic representation. We introduce two types: Boolean truth tables (logical equivalence) and extended truth tables (computational equivalence). These tables are automatically constructed to represent all possible input-output mappings of an expression. 

We present gMBA, a guided MBA deobfuscation framework that integrates semantic information to improve deobfuscation performance. Using a Transformer-based Seq2Seq model, our approach integrates truth tables to enhance internal representation learning. Specifically, we encode both the obfuscated source code and the truth tables into vector space representations and merge them to inject into the decoder architecture. To the best of our knowledge, this is the first approach leveraging semantic representations for MBA deobfuscation.

Comparison between our gMBA and previous methods on NeuReduce dataset~\cite{feng-etal-2020-neureduce} showed that our method outperforms them for both accuracy for exact matching and BLEU metrics.  

\section{Boolean and Extended Truth Tables}

MBA deobfuscation is NP-hard, meaning that no general deterministic algorithm can solve it efficiently~\cite{zhou2007information}. Previous approaches have treated MBA obfuscation as a black box without understanding its mechanism. While Seq2Seq models have been applied, they rely solely on syntactic patterns.
To incorporate semantic information, we introduce truth tables, representing an expression’s output for all input combinations, encapsulating its semantics. 

\paragraph{Truth Table Extraction for MBA Expressions}

Given Boolean variables $x_1, x_2, \dots, x_n$ where each $x_i$ takes values from $\{0,1\}$, we define a Boolean function derived from an MBA expression:
\vspace{-3pt}
\begin{equation}
    f: \{0,1\}^n \to \mathbb{Z}
\end{equation}

which maps each possible input combination to an integer output. The truth table representation of $f$ is constructed by evaluating $f(x_1, \dots, x_n)$ for all $2^n$ possible input assignments.

For a given function $f(x_1, \dots, x_n)$, we define the truth table as a vector $\mathbf{T}_f$ of length $2^n$, where each entry corresponds to the function value for a specific binary input configuration:
\begin{equation}
    \mathbf{T}_f =
    \begin{bmatrix}
        f(0,0,\dots,0) \\
        f(0,0,\dots,1) \\
        \vdots \\
        f(1,1,\dots,1)
    \end{bmatrix}.
\end{equation}

Each row of the truth table corresponds to a binary tuple $(x_1, \dots, x_n)$ sorted in lexicographical order, and the resulting vector uniquely represents the function’s output across all possible inputs. For an original expression $f(x_1, \dots, x_n)$ and an MBA-obfuscated expression $g(x_1, \dots, x_n)$, we compute their truth table vectors $\mathbf{T}_f$ and $\mathbf{T}_g$, respectively.

If $\mathbf{T}_f = \mathbf{T}_g$, then $f$ and $g$ are functionally equivalent despite their syntactic differences. This property enables MBA deobfuscation by reducing complex expressions to their canonical form. 

\section{Methods}

In this study, truth tables serve as features to enhance the model’s understanding of Mixed Boolean-Arithmetic (MBA) expressions. 
To properly ground the model’s reasoning in both syntactic and semantic evidence, we introduce a feature fusion mechanism that concatenates the encoder’s latent embedding—capturing the syntactic patterns of obfuscated expressions—with a vectorized truth table representation that preserves raw semantic evaluation values.



In this section, we describe gMBA, where both the obfuscation code and a semantic representation of the truth table serve as inputs.

\subsection{Sequence-to-sequence Architecture}
\paragraph{Encoder}
The encoder processes the input sequence $\mathbf{x}$ and generates latent representations $\mathbf{H}_L$ through stacked layers.
\begin{equation}
\begin{split}
    &\mathbf{H}_0 = \mathbf{E}_{\text{input}} \\
    &\mathbf{H}_l = \text{LN}(\mathbf{H}_{l-1} + \text{MultiHeadAttn}(\mathbf{H}_{l-1})) \\
    &\mathbf{H}_l = \text{{LN}}(\mathbf{H}_l + \text{FFN}(\mathbf{H}_l))
\end{split}
\end{equation}
where LN and FFN refers to layer normalization and feedforward networks, respectively. After passing through $L$ layers, the final encoder output $\mathbf{H}_L$ is obtained. 

\paragraph{Concatenation with Truth Table}
To integrate semantics
the truth table vector $\mathbf{T}$ is projected and concatenated with the encoder output:
\begin{equation}
\begin{split}
    &\mathbf{T} = \text{Unsqueeze}(\text{Linear}(\mathbf{T})) \\
    &\mathbf{H}_{\text{final}} = \text{Concat}(\mathbf{H}_L, \mathbf{T})
\end{split}
\end{equation}
where $\text{Linear}(\cdot)$ projects $\mathbf{T}$ into the embedding space. The concatenation position varies, and a \texttt{<sep>} token is added when explicitly separating syntax and semantics.

\paragraph{Decoder}
The decoder generates the output sequence $\mathbf{y}$ based on $\mathbf{H}_{\text{final}}$.
\begin{equation}
\begin{split}
    \mathbf{Y}_0 &= \text{Embed}(\mathbf{y}) + \text{PosEnc}(\mathbf{y}) \\
    \mathbf{Y}_l &= \text{LN}(\mathbf{Y}_{l-1} + \text{MaskedMultiHeadAttn}(\mathbf{Y}_{l-1})) \\
    \mathbf{Y}_l &= \text{LN}(\mathbf{Y}_l + \text{MultiHeadAttn}(\mathbf{Y}_l, \mathbf{H}_{\text{final}})) \\
    \mathbf{Y}_l &= \text{LN}(\mathbf{Y} + \text{FFN}(\mathbf{Y}))
\end{split}
\end{equation}

\paragraph{Output Probability Calculation}
The final decoder output is mapped to a probability distribution
\begin{equation}
    \mathbf{y}_{\text{pred}} = \text{Softmax}(\text{Linear}(\mathbf{Y}_L))
\end{equation}
where $\text{Linear}(\cdot)$ maps the decoder output to the vocabulary space, and $\text{Softmax}(\cdot)$ generates token probabilities.

\subsection{Integration of Truth Table Information}

\paragraph{Truth Tables as Semantics of MBA Expressions}

To integrate semantic information, two types of truth tables were generated: Boolean truth tables (bool tt.) and extended truth tables (extended tt.). 

The Boolean truth table evaluates MBA expressions for all input combinations, ensuring logical equivalence. In the function representation (Section 2), the range of $f$ is ${0,1}$, capturing the semantic relationship between obfuscated (src) and simplified (trg) expressions. This establishes a solid foundation for understanding logical behavior, independent of syntactic complexity.

On the other hand, the extended truth table encodes extended operations in MBA expressions, evaluating numeric results across input combinations to preserve computational semantics. In the function representation (Section 2), the range is $\mathbb{Z} \mod m$, where $m > 2$. While the Boolean truth table captures binary equivalence, the extended truth table reflects operational depth and complexity, providing a complementary perspective.

Table~\ref{tab:tt} shows the Boolean truth table and extended truth table for 
$(a\land b) + 2(a\&b)$, yielding different results.

\vspace*{-0.3cm}
\begin{table}[H]
\centering
\scriptsize
\[
\begin{array}{c  c c c c c}
\toprule
   & \multicolumn{5}{c}{\mathrm{Boolean \ Truth \ Table}} \\ 
\midrule
 a, b   & a + b & a \oplus \ b    & a \& b    & 2(a \& b) & (a\oplus b) + 2(a\&b) \\ 
\hline
 0, 0   & 0     & 0         & 0         & 0         & 0     \\ \vspace*{-2pt}
 0, 1   & 1     & 1         & 0         & 0         & 1     \\ \vspace*{-2pt}
 1, 0   & 1     & 1         & 0         & 0         & 1     \\  \vspace*{-2pt}
 1, 1   & 0     & 0         & 1         & 0         & 0     \\  
\midrule
   & \multicolumn{5}{c}{\mathrm{Extended \ Truth  \ Table}} \\ 
\midrule
 a, b   & a + b & a \oplus \ b & a \& b    & 2(a \& b) & (a\oplus b) + 2(a\&b) \\ 
\hline
 0, 0   & 0     & 0         & 0         & 0         & 0     \\\vspace*{-2pt}
 0, 1   & 1     & 1         & 0         & 0         & 1     \\\vspace*{-2pt}
 1, 0   & 1     & 1         & 0         & 0         & 1     \\\vspace*{-2pt}
 1, 1   & 2     & 0         & 1         & 2         & 2     \\  
\bottomrule
\end{array}
\]
\vspace*{-0.4cm}
\caption {Example of boolean truth table (upper) and extended truth table (lower) of $(a\oplus b) + 2(a\&b)$}
\label{tab:tt}
\end{table}

\paragraph{Integration Strategy}
We explore three methods to integrate the encoder output (\(H\)) with the truth table (\(T\)): addition, token-level concatenation, and hidden-dimension concatenation. These strategies are defined in Equation~\ref{eqn:integration_strategy}.

\vspace*{-0.4cm}
\begin{equation}
\text{Integrate}(H, T) = 
\begin{cases} 
    H + T       & \textit{add} \\ 
    [H; T]      & \textit{tok-level cat} \\ 
    H \oplus T  & \textit{hid-dim cat}
\end{cases}
\label{eqn:integration_strategy}
\end{equation}

\textbf{Addition}: The truth table (\(T\)) is repeated along the token dimension and added element-wise to the encoder output (\(H\)), preserving its shape \(B \times T \times D\). For clarity, \( \mathbf{T} \) is broadcast (repeated) across the token dimension, ensuring it matches the shape of \( \mathbf{H} \) before the element‐wise addition.

\vspace*{-0.17cm}
\begin{equation}
H'_{i,j,k} = H_{i,j,k} + T_{i,k}    
\label{eqn:add}
\end{equation}

\textbf{Token-level Concatenation}: The truth table (\(T\)) is reshaped to \(B \times 1 \times D\) and appended along the token dimension, resulting in \(H' \in \mathbb{R}^{B \times (T+1) \times D}\).

\vspace*{-0.17cm}
\begin{equation}
H' = [H; T]    
\label{eqn:tok_level_cat}
\end{equation}

\textbf{Hidden-dimension Concatenation}: \(T\) is repeated along the token dimension and concatenated along the hidden dimension, forming \(H' \in \mathbb{R}^{B \times T \times (D + D')}\).

\vspace*{-0.2cm}
\begin{equation}
H' = H \oplus T    
\label{eqn:hiddim_cat}
\end{equation}

\paragraph{Concatenation Strategy}

To optimize the concatenation strategy, we experimented with various configurations, focusing on token-level concatenation, which effectively incorporates truth table information. The tested strategies, with key variables are summarized as follows: (1) \textit{Concat Position}: the order in which semantic (truth table) and syntactic (encoder output) information are combined, (2) \textit{Seperator Token}: whether a special separator token is inserted to distinguish between syntax and semantics, and (3) \textit{Semantics}: the type of semantic information used, including Boolean truth table, extended truth table, or both.

\subsection{Overall Architecture}

\begin{figure}[htb!]
    \centering
    \includegraphics[trim={200} {50} {200} {50} cm, clip, width=\linewidth]{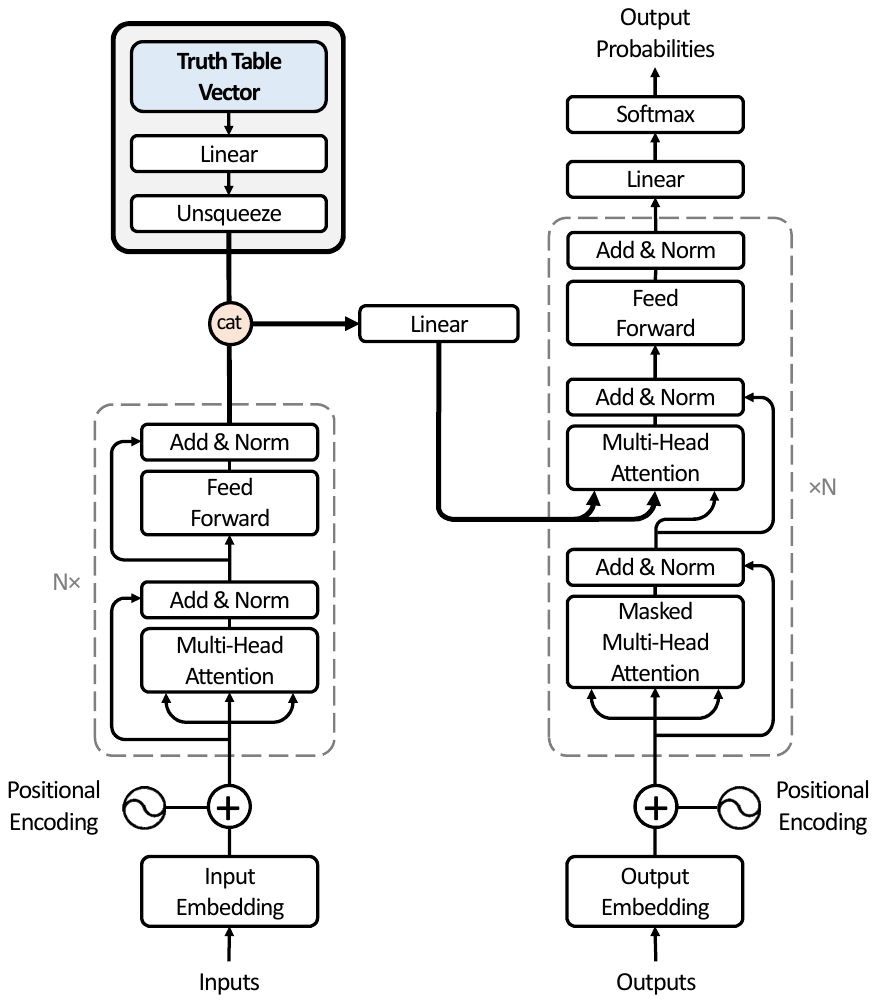}
    \caption{Model Architecture}
    \label{fig:model-arch}
\end{figure}

Figure~\ref{fig:model-arch} illustrates our model architecture, which extends the standard Transformer by integrating a truth table vector to enhance semantic understanding of mathematical expressions. The encoder follows the standard multi-head self-attention and feed-forward layers, while the truth table vector is processed through a linear transformation and concatenated with the encoder representation. This augmentation provides explicit semantic guidance to the model.

The decoder incorporates masked self-attention and attends to both the encoder output and the augmented representation, enabling it to better differentiate semantically similar expressions. This architecture improves the model’s ability to recover precise mathematical expressions during deobfuscation.

\section{Experiments}

\subsection{Experimental Settings} \label{subsec:experiment_setting}

\paragraph{Dataset} We used \texttt{NeuReduce} dataset for our experiments, using the same settings as in previous work~\cite{feng2020neureduce}. Based on standard split, we divided the dataset into 80k, 20k, and 10k instances for training, validation, and test datasets, respectively.
Appendix~\ref{app:data_stats} provides further statistics. 

\paragraph{Implementation Details} The model's performance was evaluated using two metrics: exact matching accuracy and the BLEU score for n-gram matching. We used greedy decoding in all experiments. We trained the model with hidden dimension size 256 and max length 104. 
The baseline model is a vanilla transformer network used in the previous work that does not leverage truth tables.

\subsection{Results}
The experimental results on \texttt{NeuReduce} dataset are shown in Table~\ref{tab:main}.
As shown in the table, gMBA consistently achieves higher accuracy and BLEU scores compared to the Vanilla model across all the configurations. This improvement highlights the advantage of leveraging the semantics of MBA expressions, enabling the correct deobfuscation of MBA expressions with similar syntax but different evaluation results. Specifically, gMBA outperforms the Vanilla model, achieving an accuracy improvement of at least 21.72\% and up to 51.94\%, along with a BLEU score increase of at least 8.78 and up to 17.49 points; the extended truth table with the <sep> token, following syntactic information, yields the highest accuracy of 92.78\%.

\begin{table}[H]
\centering
\begin{adjustbox}{width=\columnwidth,center}
\begin{tabular}{llccc}
\toprule
\textbf{Model} & \textbf{Concat Position} & \textbf{<sep>} & \textbf{Acc (\%)} & \textbf{BLEU (\%)} \\
\midrule
Vanilla & - & - & 40.84  & 78.05  \\
\hline
 
\multirow{6}{*}{\textbf{bool tt.}} 
 & \multirow{2}{*}{back} & N      & 65.05  & 87.84  \\
 &  & Y & 65.17  & 87.28  \\
 \cline{2-5}
 & \multirow{2}{*}{front} & N      & 66.01  & 87.86  \\
 &  & Y & 63.91  & 87.17  \\
 \cline{2-5}
 & \multirow{2}{*}{back, front of <pad>} & N & 62.56 & 86.83 \\
 &  & Y & 65.99 & 87.77 \\
\hline

\multirow{6}{*}{\textbf{extended tt.}} 
 & \multirow{2}{*}{back} & N      & 91.69  & 94.96  \\
 &  & Y & 92.59  & 95.39  \\
 \cline{2-5}
 & \multirow{2}{*}{front} & N      & 92.06  & 94.99  \\
 &  & Y & 87.69  & 93.97  \\
 \cline{2-5}
 & \multirow{2}{*}{back, front of <pad>} & N & 91.71 & 95.00 \\
 &  & Y & \textbf{92.78}  & \textbf{95.53} \\
\hline

\multirow{6}{*}{\textbf{both}} 
 & \multirow{2}{*}{back} & N      & 90.01  & 95.14  \\
 &  & Y & 88.83  & 94.83  \\
 \cline{2-5}
 & \multirow{2}{*}{front}     & N      & 91.01  & 95.48  \\
 &  & Y & 89.19  & 94.98  \\
 \cline{2-5}
 & \multirow{2}{*}{back, front of <pad>} & N & 91.57 & 95.51 \\
 &  & Y & 88.03 & 94.49 \\
\bottomrule
\end{tabular}
\end{adjustbox}
\caption{Impact of different truth table integration strategies on deobfuscation performance.
Vanilla refers to the baseline~\cite{feng2020neureduce} Transformer model without truth table information.
}
\label{tab:main}
\end{table}

\subsection{Analysis}

\paragraph{Impact of Truth Tables on Accuracy and Semantic Understanding}
In Table~\ref{tab:main}, we observe that the vanilla model achieves a relatively high BLEU score (78\%) but a much lower exact-match accuracy (40\%). This discrepancy suggests that while the model can generate expressions that look similar to the reference, it often fails to produce the exact target expression. In contrast, when we introduce truth table strategies, both BLEU and accuracy improve, and—crucially—the gap between these two metrics narrows. We interpret this as evidence that the added semantic information (via truth tables) helps the model better distinguish among closely related expressions. Moving from the vanilla (no semantic) setup to Boolean truth tables and then arithmetic truth tables, the model gains deeper semantic understanding, allowing it to 
shift from superficially similar expressions to precisely aligned original expressions. Furthermore, these results highlight the potential of neural models to learn and reason over the semantics of mathematical expressions, capturing underlying structures beyond mere syntactic similarity.

\paragraph{Effectiveness of the Extended Truth Tables}

In our observations, one of the key challenges in MBA deobfuscation lies in accurately generating the constants and coefficients within an expression. While Boolean truth tables provide only binary outcomes, the extended truth table encodes richer numerical information—specifically, multi-bit evaluation results—which can assist the model in identifying and reasoning about underlying arithmetic patterns.

This additional numerical signal enables the model to more effectively infer structural and semantic properties of the expression, including magnitude and relational cues that are not readily available from binary outputs alone. We believe that the intermediate numerical guidance provided by the extended truth table significantly enhances the learning process, ultimately leading to better deobfuscation performance.

\paragraph{Comparison with Fine-Tuned PLMs and LLMs}

One potential concern is whether pre-trained language models and large language models, with general mathematical and logical knowledge, can outperform specialized models on MBA tasks.

We additionally conducted experiments and the results are shown in Table~\ref{tab:plm_comparison}.
For comparison, we selected BART\footnote{https://huggingface.co/docs/transformers/ko/model\_doc/bart}~\cite{lewis2019bart} and T5\footnote{https://huggingface.co/docs/transformers/model\_doc/t5}~\cite{raffel2020exploring} as encoder-decoder architectured  PLMs and recent LLaMA models\footnote{https://huggingface.co/meta-llama} for LLMs.
We fine-tuned and evaluated them with the same data described in \ref{subsec:experiment_setting}.

\begin{table}[H]
\centering
\begin{adjustbox}{width=0.8\columnwidth,center}
\begin{tabular}{lcc}
\toprule
\makecell[c]{\textbf{Model}} & \multicolumn{2}{c}{\textbf{Metrics (\%)}} \\
& Acc & BLEU \\
\midrule
NeuReduce \tablefootnote{The baseline was reimplemented based on \cite{feng2020neureduce}.} & 40.84   & 78.05 \\
\midrule
LoRA Fine-tuned BART  & 47.96   & 81.91 \\
Fine-tuned BART       & 52.12   & 74.36 \\
LoRA Fine-tuned T5    & 50.90   & 82.30 \\
Fine-tuned T5         & 54.50   & 83.91 \\
\midrule
LoRA Fine-tuned LLaMA 3.2 3B & 37.29 & 75.13 \\
LoRA Fine-tuned LLaMA 3.1 8B & 38.56 & 79.17 \\
\midrule
gMBA (\textit{proposed}) & \textbf{92.78} & \textbf{95.53} \\
\bottomrule
\end{tabular}
\end{adjustbox}
\caption{Performance comparison of fine-tuned PLMs, instruction-tuned LLMs and the proposed method on MBA deobfuscation using the dataset of NeuReduce}
\label{tab:plm_comparison}
\end{table}

Table~\ref{tab:plm_comparison}'s findings suggest that the prior mathematical and logical knowledge embedded in PLMs alone is insufficient for effective MBA deobfuscation, highlighting the necessity of specialized architectures that explicitly capture the semantics of MBA expressions. 
It also shows the limitation of the LLMs and we anlayzed this problem is attributed to the pretraining objective of LLMs, which optimizes for generating fluent natural language. The underlying token distribution is shaped by linguistic plausibility rather than the syntactic precision required for symbolic tasks. Consequently, even with instruction-tuning, these models struggle to meet the strict accuracy demands of MBA deobfuscation, unlike our proposed method, gMBA.

\section{Conclusions}
We showed that incorporating truth tables derived from semantic information is crucial for MBA deobfuscation. To incorporate truth tables, we proposed a gMBA that leverages a Transformer-based Seq2Seq framework. Our analysis showed that gMBA outperforms both traditional methods and a naive Transformer network, resulting in improving performance.

\section*{Limitations}

In terms of deobfuscation accuracy, gMBA based on neural networks achieves an accuracy in the 90\% range, whereas the state-of-the-art (SOTA) logic-based methods achieve 100\% accuracy \cite{reichenwallner_simplification_2023}. Despite this limitation, gMBA has potential for improvement and scalability through data augmentation, enhancing its ability to handle diverse obfuscation patterns, unlike highly accurate logic-based approaches that struggle with scalability due to the complexity of crafting rules.

\section*{Acknowledgements}

This work was supported by the National Research Foundation of Korea(NRF) grant funded by the Korea government(MSIT)(RS-2024-00456953). \newline
This work was supported by Institute of Information \& communications Technology Planning \& Evaluation (IITP) grant funded by the Korea government(MSIT) (RS-2022-00143911, AI Excellence Global Innovative Leader Education Program)

\bibliography{custom, anthology}

\appendix

\section{Dataset Statistics} \label{app:data_stats}

Table~\ref{tab:dataset_stats} provides the detailed statistics of the dataset used in our experiments. The dataset consists of expressions with varying numbers of variables, operators, and lengths. The Train-small dataset contains 100K samples, while the Test set consists of 10K samples. 
The number of variables and operations follows a wide distribution, reflecting the complexity of mathematical expressions encountered during training and evaluation. This distribution ensures that the model is tested on expressions of varying difficulty, supporting a robust assessment of its generalization ability.

\begin{table}[h!]
    \renewcommand{\arraystretch}{1.1}
    \centering
    \resizebox{\columnwidth}{!}{
        \begin{tabular}{ccccc}
        \toprule
        \textbf{} & \multicolumn{2}{c}{\textbf{Train-small}} & \multicolumn{2}{c}{\textbf{Test}} \\
        \textbf{} & \texttt{src} & \texttt{trg} & \texttt{src} & \texttt{trg} \\
        \midrule 
        \textbf{Size} & 100K & 100K & 10K & 10K \\
        \textbf{\# of Vars} & \(17.0 \pm 15.0\) & \(4.5 \pm 3.5\) & \(16.5 \pm 14.5\) & \(4.5 \pm     3.5\) \\
        \textbf{\# of Ops} & \(26.0 \pm 23.0\) & \(6.0 \pm 6.0\) & \(25.0 \pm 23.0\) & \(6.0 \pm 6.0\) \\
        \textbf{Length} & \(54.0 \pm 46.0\) & \(18.0 \pm 17.0\) & \(52.0 \pm 48.0\) & \(18.0 \pm 17.0\) \\
        \bottomrule
        \end{tabular}
    }
    \caption{Statistics of the Dataset (excluding Train-large)}
    \label{tab:dataset_stats}
\end{table}

\section{Engineering}

\begin{table}[htb!]
\small
    \centering
    \begin{tabular}{lccc}
        \toprule
        \textbf{Integration Strat.} & \textbf{Ver.} & \multicolumn{2}{c}{\textbf{Metrics (\%)}} \\
        & & \textbf{Acc} & \textbf{BLEU} \\
        \midrule

        
        Addition & - & 64.29 & 87.69 \\
        \hline
        \multirow{4}{*}{Token-level Concat}  & v1 & 58.60 & 84.18 \\
        & v2 & 63.87 & 87.31 \\
        & v3 & \textbf{65.05} & \textbf{87.84} \\
        & v4 & 35.77 & 74.11 \\
        & v5 & 61.89 & 85.91 \\
        & v6 & 61.95 & 86.62 \\
        & v7 & 61.82 & 86.01 \\
        
        \hline
        \multirow{7}{*}{Hidden-dim Concat} & v1 & 0.44 & 34.56 \\
        & v2 & 0.06 & 17.88 \\
        & v3 & 0.14 & 19.81 \\
        & v4 & 0.26 & 23.32 \\
        & v5 & 0.12 & 15.26\\
        & v6 & 0.13 & 18.55 \\
        & v7 & 0.13 & 16.99 \\

        \bottomrule
    \end{tabular}
    \caption{Comparison of Accuracy and BLEU Metrics for Truth Table Integration Techniques to Assess Syntax-Semantics Fusion Methods in Transformer-based Deobfuscation}
    \label{tab:integration_strategy-appendix}
\end{table}

To maximize the performance of each integration strategy, we conducted extensive engineering experiments to determine the optimal placement and type of normalization layers. Using the Boolean truth table as a test case, we explored various configurations within the concatenation-based integration methods to identify settings that yield the best accuracy and BLEU scores. Table~\ref{tab:integration_strategy-appendix} summarizes the results, highlighting the effect of different normalization strategies. These insights helped in refining our final implementation, ensuring that each approach was optimized for its best possible performance.

When using the hidden-dimension concatenation strategy, we observed that the matrices encoding the expression’s syntax tend to collapse, causing a marked drop in performance. Hence, we concluded that this issue cannot be overcome by engineering efforts alone.

\section{Analysis of Errors}

\begin{figure}[h] \small
\begin{verbatim}
  trg1  : -2*(x|~y)-(~x)-1
  pred1 : -2*(x|~y)-(~(~x)
  trg2  : -3*(x&~y)-4*(~x&y)
  pred2 : -3*(x&~)-4*(~x&y)
  trg3  : -17*(x|y|z)
  pred3 : -16*(x|y|z)
  trg4  : x&y
  pred4 : x-(((((((((((
\end{verbatim}
\label{fig:error_examples} 
\caption{Samples of misprediction}
\end{figure}

In reviewing the model's mispredicted expressions, we identified several recurring error patterns that highlight the system's current limitations. In most cases, the answers started with the same subexpression but ended incorrectly. There were also many cases that had generated almost identical to the correct answer but with a slightly different constant or operator. We observed a few instances, wherein the model introduces extraneous characters or repeated segments that deviate from the expected structure.

\end{document}